\documentclass{article}
\usepackage{multicol}
\usepackage{amsmath}
\usepackage{amsfonts}
\usepackage{amssymb}
\usepackage{amsthm}

\begin{document}
\title{%
\normalsize \bfseries QKD Quantum Channel Authentication}

\author{\small J.T. Kosloski
%}
 \\%\affiliation{
 \emph{\footnotesize
National Security Agency, 9800 Savage Road, Fort George G. Meade, Maryland 20755}
\thanks{jtkosloski@yahoo.com}}
%\email{jtkosloski@yahoo.com}
%\pacs{03.67.Dd} %Quantum Crypt
%\pacs{03.67.-a} %Quantum infor
\date{(\footnotesize Dated: \today)}

\maketitle \vspace{-1cm}
\begin{abstract}Several simple yet secure protocols to authenticate the quantum channel of
various QKD schemes, by coupling the photon sender's knowledge of a shared secret and the
QBER Bob observes,  are presented. It is shown that Alice can encrypt certain portions of
the information needed for the QKD protocols, using a sequence whose security is based on
computational-complexity, without compromising all of the sequence's entropy. It is then
shown that after a Man-in-the-Middle attack on the quantum and classical channels, there is
still enough entropy left in the sequence for Bob to detect the presence of Eve by
monitoring the QBER.
 Finally, it is shown that the principles presented can be implemented to authenticate the quantum channel associated with any
type of QKD scheme, and they can also be used for Alice to authenticate Bob.
\end{abstract}

\begin{multicols}{2}

\section{Introduction}
Quantum Key Distribution (QKD) has gained considerable interest in the academic and
commercial sectors in recent years because of its ability to offer absolute security against
all attacks that can be carried out on classical and quantum computers. This is in stark
contrast to current classical public-key schemes that have been shown to be vulnerable to
attacks on a quantum computer \cite{shor}.  However, these same classical schemes do have a
significant advantage in that they can be used to authenticate messages and eliminate
Man-in-the-Middle attacks, at least when Eve (the adversary) is limited to a classical
computer. In the absence of an authenticated public channel, most QKD protocols, such as
BB84 \cite{BB84}, are not secure against Man-in-the-Middle attacks.
%Most QKD protocols, such as BB84 \cite {BB84}, are not secure in the classical or
%the quantum sense against Man-in-the-Middle attacks, so QKD protocols alone do not provide
%unconditional security.

The current method to secure commercially viable QKD protocols against such an attack is to
authenticate the classical communications between Alice and Bob.  This prevents Eve from
establishing key with either one because she would not be able to carry out the classical
communications necessary for the protocols, and she would be limited to attacks that
increase Bob's observed quantum bit error rate (QBER).  The Wegman-Carter authentication
scheme \cite{w-c_authen} and variations thereof \cite{zielinger_authen} seem to be the most
commonly implemented methods to authenticate QKD public channels.  They also seem to be
sufficient to protect against Man-in-the-Middle attacks.  However, these schemes do not
actually authenticate the users of a quantum channel, and there could be situations where
this is desired.

There have been several quantum authentication protocols developed for the purpose of
authenticating quantum messages \cite{quant_authen_1} \cite{quant_authen_2}
\cite{quant_authen_3}, with much of the focus being on the use of entanglement.  A quantum
message is a normal message sent over a quantum channel using quantum codes. On the other
hand, only random bits are transmitted over the quantum channel in QKD, and all messages are
sent over the classical channel. In many of the quantum message authentication schemes, a
shared secret is used to encrypt a message that is transmitted using one of several quantum
codes. An imposter is then detected by monitoring the errors in the code words. One problem
these schemes have is the inherent structure in the codes and Eve's ability to take
advantage of possible correlations between two sequential bits, resulting from the
structures of quantum codes.
%prior to the code words being encrypted.
However, in QKD, there are no bit-to-bit correlations, assuming a perfectly random raw bit
sequence, so it seems reasonable that QKD could be simpler to authenticate than a quantum
message.
%Ideally,
%quantum channel authentication should rely on available technology that can be implemented
%with currently available QKD systems and it should require minimal computational and
%infrastructural overhead.
In this article, it is shown that the quantum-based %absolute
security of entanglement-based authentication may not be necessary, and that
computational-complexity-based schemes are sufficient to authenticate the quantum channel of
a QKD system.

Four protocols are presented, each of which requires only a shared secret and a
key-expansion function, in addition to
%, and then utilizes
the standard QKD protocols, to detect an imposter. Through examples of Man-in-the-Middle
attacks, it is shown that even though information about the shared secret will be leaked to
Eve during a QKD session, as long as determining the shared secret (given the expanded key)
requires more computation than is possible in a few seconds, there is enough entropy
remaining in the expanded key for Bob to detect the presence of an imposter by monitoring
the QBER. Finally, it is shown that the basic principles used for these protocols can be
implemented to authenticate the quantum channel associated with any type of QKD scheme, and
that these protocols can also be used for Alice to authenticate Bob.

%Several simple yet secure protocols to authenticate the quantum channel of various QKD
%schemes are presented.  It is shown that Alice can encrypt certain portions of the
%information needed for the QKD protocols, such as basis choices or bit choices, using a
%sequence whose security is based on computational-complexity, without revealing the entire
%sequence.  It is then shown that after a man-in-the-middle attack on the quantum and
%classical channels there is still enough entropy left in the sequence for Bob to detect the
%presence of an imposter by monitoring the QBER.

\section{Protocols}Consider the situation where Alice and Bob are going to generate key
using BB84 and have an $n$-bit shared secret $K$, where $n$ is chosen based on the level of
desired security.  Also suppose that Alice and Bob have agreed on a key expansion function
$F()$, which need not be kept secret, that they consider secure from time-limited
cryptographic attacks on both classical and quantum computers.  The time it takes to
determine $K$ given $F(K)$ needs to be longer than the time it takes to perform a QKD
session.

For notational purposes, let $F(K)^i$ be the $i^{th}$ bit of the expanded sequence, which is
synchronized with clocks at Alice and Bob.  Let $x^t$ and $y^t$ be Alice's bit and basis
choice at time $t$, and let $z^t$ be Bob's basis choice.  Let the observable being used
(phase, polarization, orbital angular momentum, ...) be represented by $\Gamma$, and let the
two conjugate bases be denoted by $\Gamma_0$ and $\Gamma_1$.  To put the notation into
context, the quantum portion of BB84 is carried out when Alice sends $\Gamma_{y^t}=x^t$ and
Bob measures $\Gamma_{z^t}$.

Each of the protocols below allow Bob, after Error Correction (EC), to conclude whether or
not the photons originated with an impersonator, as well as whether or not he communicated
with an impersonator during either sifting or EC, depending on the protocol.  That is not to
say that these protocols protect against the possibility that Eve is intercepting
information, which is the purpose of the actual QKD protocol, but it does say that the
information did not originate with Eve.

 Note that $F(K)$, which is a pseudo-random bit sequence, will not be
available to Eve for analysis until she has recovered the random bit stream with which it is
combined, such as Alice's bit or basis choices, because a random stream $Xored$ with any
stream produces a random stream. So, Eve will not even be able to begin working on the
recovery of $K$ until after EC. It should also be noted that in the protocols below, $1a$
and $2a$ have timing limitations that $1b$ and $2b$ do not. Namely, $1a$ and $2a$ are only
secure if Eve does not have the opportunity to complete the entire Alice-Eve session before
starting Eve-Bob because Eve can simply omit sending a photon to Bob for the times that
correspond to $j\in\{t\}$ for which he did not learn $F(K)^j$, and Bob would attribute a
lack of detection events to attenuation of the photons. Conversely, $1b$ and $2b$ force Eve
to use a continuous stream of $F(K)$ starting at $F(K)^0$, so she cannot avoid times for
which she does not know $F(K)^t$. Also note that the timing requirement for the first two
protocols is not unreasonable and can easily be met.
%  For the protocols presented in this article it is not
%necessary for Alice and Bob to keep their choice of key expansion functions a secret, but
%there is also no reason to release the information.
%
\subsection*{Protocol 1.a}
%\large{$Protocol\text{ }1.a$}\normalsize
%
\begin{enumerate}

\item
\begin{enumerate}
\item Alice sends a photon with $\Gamma_{y^t}=x^t \oplus F(K)^t$.

\item Bob measures $\Gamma_{z^t}$, and records\\ $x'^t=\Gamma_{z^t}\oplus F(K)^t$.
\item This step continues until enough photons have been sent for Bob to accurately calculate
the QBER.
\end{enumerate}

\item Alice and Bob perform bit distillation.  Alice publicly discloses the set of her basis
choices, $\{y\}$.  Bob then compares $\{y\}$ to $\{z\}$ and publicly discloses a list of the
times that have valid bits.  (Alice $\rightarrow$ Bob Sifting using $\{y\}$ and $\{z\}$)

\item Alice and Bob perform EC on the bits of $\{x\}$ and $\{x'\}$ retained after sifting,
 using some agreed-upon scheme such as CASCADE \cite{cascade}.  There is a possibility that
the error correction scheme used does not correct all of the errors, but corrects for some
maximum error rate, $\Delta$, with a high degree of certainty.  $\Delta$ could either be a
limitation of the correction scheme or Alice's unwillingness to correct more than a certain
number of errors.  For simplicity, suppose that $QBER \le \Delta$ implies there will be no
errors left after EC (with some degree of certainty) and $QBER>\Delta$ implies there will be
about a $(QBER - \Delta )$ error rate after EC.

\item Bob makes a conclusion about the security of the error-corrected bits.  If the QBER is
too high, Bob concludes that either Eve has gained too much information concerning the key
that he established with Alice (standard BB84 conclusion) or that Alice did not send the
original photons (conclusion concerning the authenticity of the photons).

\item Either Alice and Bob perform privacy amplification to create final keys, or they start
over.

\item Alice and Bob create a new K. Alice and Bob take $n$ secure bits, either pre-placed or established during a QKD
session, and create a new $K$ to authenticate the next QKD session.\\

\end{enumerate}
\subsection*{Protocol 1.b}
%\large{$Protocol\text{ }1.b$}\normalsize

\begin {enumerate}
\item
\begin {enumerate}
\item Alice sends a photon with $\Gamma_{y^t}=x^t$.

\item Bob measures $\Gamma_{z^t}$, and records  $x'^t=\Gamma_{z^t}$.

\item This step continues until enough photons have been sent for Bob to accurately calculate
the QBER.
\end{enumerate}

\item Alice $\rightarrow$ Bob Sifting using $\{y\}$ and $\{z\}$.  Bob and Alice then apply the stream $F(K)$ to the
bits retained after sifting with a bit-wise Xor.

\item Alice and Bob perform EC on the bits of $F(K)$ applied to the bits of $\{x\}$ and $\{x'\}$
retained after sifting.

\item Bob makes a conclusion about the security of the error-corrected bits.

\item Either Alice and Bob perform privacy amplification to create final keys, or they start
over.

\item Alice and Bob create a new $K$.

\end {enumerate}
\subsection*{Security of Protocol 1}
The QBER for these protocols is a function of $\Gamma$ being measured and any tampering that
may occur on the quantum channel as well as the original sender's knowledge of the $F(K)$
sequence. If the established QBER is sufficiently low, Bob concludes that the person he is
communicating with for the EC, over the classical channel, either knows $(\Gamma_{y^t}$ and
$F(K)^t)$ or $(x^t$ and $F(K)^t)$, for the half of the time slots that correspond to his
detection events. This doesn't directly guarantee that the photon was originally sent by
Alice, but rather guarantees that the person Bob is communicating with for the EC, over the
classical channel, has knowledge that only the sender of the photons would have as well as
knowledge of $F(K)$, which only Alice has.  Put another way, this protocol guarantees that
Bob is communicating classically with the sender of the photons for the EC, and that the
sender knows $F(K)$. Therefore, the
original sender must be Alice.  %Note that through an interactive EC Alice will also be able
%to verify, to some extent, the identity of Bob because she will be able to observe roughly
%how many errors he is correcting.

To understand the security of these protocols, consider the following Man-in-the-Middle
attack against Protocol 1.a, assuming the timing restrictions for Protocol 1.a, as noted
above, have been met.% (similar security when carried out against 1.b, but without the timing
%restrictions)

\begin{enumerate}
\item
\begin{enumerate}
\item Alice sends a photon with $\Gamma_{y^t}=x^t \oplus F(K)^t$.

\item Eve measures $\Gamma_{\mu^t}$, and records $\chi^t=\Gamma_{\mu^t}.$

\item Eve sends a photon with $\Psi_{\nu^t}=\xi^t$, where $\Psi$ and $\Gamma$ are the same
observable.

\item Bob measures $\Psi_{z^t}$, and records \\$x^t=\Psi_{z^t}\oplus F(K)^t$.

\item This step continues until enough photons have been sent for Bob to accurately calculate
the QBER.
\end{enumerate}

\item Alice and Eve perform bit distillation. Alice sends Eve the set of her basis choices,
$\{y\}$. Eve then compares $\{y\}$ to $\{\mu\}$ and sends Alice a list of which bits to
include, with about half of them being discarded. (Alice $\rightarrow$ Eve sifting)

\item Alice and Eve perform EC. Eve didn't know $F(K)$, so she has a $.50$ error rate in
her key, relative to Alice. After EC, Eve still has a $\alpha = max\{0,(.50-\Delta) \} $
error rate for the bits retained after sifting. If $\Delta$ is sufficiently small, this
could prevent Eve from establishing perfect keys with Alice, and could allow Alice to detect
an imposter while Alice and Eve are communicating with the keys.

After EC, Eve's $F(K)$, for the bits corresponding to events retained after sifting, has an
error rate of $\alpha$, and her error rate for the complete $F(K)$ is then $(.25
+\frac{\alpha}{2})$.

\item Eve and Bob perform bit distillation.  Eve sends Bob the set of her basis choices,
$\{\nu\}$.  Bob then compares $\{\nu\}$ to $\{z\}$ and sends Eve a list of which bits to
include, with about half of them being discarded (Eve $\rightarrow$ Bob Sifting).  As long
as Eve did not know $\{y\}$ prior to (Alice $\rightarrow$ Eve Sifting) and did not know
$\{z\}$ prior to (Eve $\rightarrow$ Bob Sifting), about half of the bits retained by Bob and
Eve will correspond to bits retained by Eve and Alice.

\item Eve and Bob perform EC.  Eve's total error rate of $F(K)$ is $(.25+\frac{\alpha}{2})$,
so her key will have a $(.25+\frac{\alpha}{2})$ error rate relative to Bob's key.

\item Bob will calculate a $.25  \le (QBER = .25+\frac{\alpha}{2}) \le .50 $ and conclude
that Eve must be involved.
\end{enumerate}

An analogous Man-in-the-Middle attack carried out against 1.b would have similar results in
practice, but without the timing restriction. Against 1.b the attack would, in theory,
induce a $QBER=\alpha$ which implies $0\le QBER \le .5$. However, keeping in mind that the
most trivial attacks against QKD produce a $QBER=.25$, it is unlikely that Alice would allow
$\Delta
> .25$ and therefore, in practice, Bob will also calculate $.25 < (QBER = \alpha) \le .5$
with protocol 1.b. Also note that during attacks on 1.a and 1.b, through interactive EC with
Bob, Eve can take advantage of some of the information she gains during the interaction to
ensure that the QBER appears to be a little lower than it actually is. This threat can be
eliminated by using forward error correction, during which no information is leaked by Bob
back to Eve.
%This threat could be reduced if the interactive error correction begins with
%the sender of the photons, presumably Alice (or Eve), sending Bob a hash of her bits, so she
%is committed, to some degree, to the bit values.  A better solution would be to use forward
%error correction; then no information is leaked by Bob back to Eve.
%
\subsection*{Protocol 2.a}
%\large{$Protocol\text{ } 2.a$}\normalsize
%
\begin{enumerate}

\item
\begin{enumerate}
\item Alice sends a photon with $\Gamma_{y^t\oplus F(K)^t}=x^t$.

\item Bob measures $\Gamma_{z^t\oplus F(K)^t}$, and records \\$x'^t=\Gamma_{z^t\oplus
F(K)^t}$.
\item This step continues until enough photons have been sent for Bob to accurately calculate
the QBER.
\end{enumerate}

\item Alice $\rightarrow$ Bob Sifting using $\{y\}$ and $\{z\}$

\item Alice and Bob perform EC on the bits of $\{x\}$ and $\{x'\}$ retained after sifting.

\item Bob makes a conclusion about the security of the error-corrected bits.

\item Either Alice and Bob perform privacy amplification, or they start over.

\item Alice and Bob create a new $K$.

\end{enumerate}
\subsection*{Protocol 2.b}
%\large{$Protocol\text{ } 2.b$}\normalsize
%
\begin{enumerate}
\item
\begin{enumerate}
\item Alice sends a photon with $\Gamma_{y^t}=x^t$.
\item Bob measures $\Gamma_{z^t}$, and records $x'^t=\Gamma_{z^t}$.

\item This step continues until enough photons have been sent for Bob to accurately calculate
the QBER.

\end{enumerate}

\item Bob publicly discloses a list of the times for which he had a detection event. Alice and
Bob remove their basis choices for times that do not correspond to detection events to
create the sets $\{y'\}$ and $\{z'\}$ respectively. Alice $\rightarrow$ Bob Sifting using
$\{y'\}\oplus F(K)$ and $\{z'\}\oplus F(K)$.%Alice publicly discloses $\{y\} \oplus F(K)$
%for sifting. Bob then compares $\{y\}$ to $\{z\}$ and publicly discloses a list of the times
%that have valid bits.

\item Alice and Bob perform EC on the bits of $\{x\}$ and $\{x'\}$ retained after sifting.

\item Bob makes a conclusion about the security of the error-corrected bits.

\item Either Alice and Bob perform privacy amplification, or they start over.

\item Alice and Bob create a new $K$.

\end{enumerate}
\subsection*{Security of Protocol 2}
 These protocols offer similar assurances to Bob
as Protocols 1.a and 1.b, except that they guarantee, after EC, that the person with which
he performed sifting is someone that knows information that only the sender of the photon
and Alice could know. In particular, after EC, Bob knows that he performed the sifting with
someone who knew both $\{y\}$ and $F(K)$, otherwise, he would have randomly selected which
bits to use for the EC and would have a substantial error rate. Therefore, the original
sender must be Alice.

The QBER is a function of $\Gamma$ being measured and any tampering that may occur on the
quantum channel, in addition to the original sender not knowing the correct $F(K)$ sequence
that was Xored to Alices's basis stream. When Bob is trying to perform EC with a user that
does not know $F(K)$, the error rate will be inflated because Bob would have randomly
selected his bits from all of the bits, roughly half of which are in the wrong basis.  Note
that, unlike in Protocol 1, the knowledge Eve can gain during interactive EC will not help
her reduce the QBER induced by her not knowing the correct basis during the sifting. So, for
protocols $2.a$ and $2.b$, Eve does not gain an advantage by performing interactive EC with
Bob as opposed to forward EC.

To understand the security of these two protocols, consider the following Man-in-the-Middle
attack against Protocol 2.a, assuming that the timing restrictions for Protocol 2.a, as
noted above, have been met (similar security when carried out against Protocol 2.b, but
without the timing restrictions).

Allow for the possibility that Alice-Eve EC is completed after Eve-Bob photon transmission,
but before Eve-Bob sifting.

\begin{enumerate}
\item Eve creates a set of times $\{\tau\}$ that correspond to bits of $F(K)$ she intends to
learn.

\item
\begin{enumerate}
\item Alice sends a photon with $\Gamma_{y^t\oplus F(K)^t}=x^t$.

\item Eve measures $\Gamma_{\mu^t}$, and records $\chi^t=\Gamma_{\mu^t}$.

\item Eve sends a photon with $\Psi_{\nu^t}=\xi^t$ if $t\in \{\tau\}$, where $\Psi$ and
$\Gamma$ are the same observable.

\item Bob measures $\Psi_{z^t\oplus F(K)^t}$, and records \\$x'^t=\Psi_{z^t\oplus F(K)^t}$
if $t\in \{\tau\}$.

\item This step continues until enough bits have been sent for Bob to accurately calculate
the QBER.
\end{enumerate}

\item Alice and Eve perform bit distillation.  Alice sends Eve the set $\{y\}$.
Eve tells Alice that they agreed on the basis selection for the times $t \in \{\tau\}$

\item Alice and Eve perform EC.  Eve didn't know $F(K)$, so she has a $.25$ error rate in
her key, relative to Alice.  After EC, Eve still has a $( .25-\Delta)$ error rate for the
bits retained after sifting.  Again, $\Delta$ could be chosen to prevent Eve from
establishing perfect keys with Alice, and could allow Alice to detect the imposter while
Alice and Eve are communicating with the keys as input to their encryption systems.

To understand what Eve knows after EC with Alice, consider the fact that Eve knows $y^t$ and
$\mu^t$ for all $t$.  Through EC she learns $y^t\oplus F(K)^t \not= \mu^t$, $t\in \{\tau\}$,
for some number of errors, which is sufficient to calculate $F(K)^t$ for these times.  For
the times that she had the correct bit value, Eve doesn't know if $y^t\oplus F(K)^t =\mu^t$
or if $y^t \oplus F(K)^t \not=\mu^t$. Therefore, Eve's copy of $\{F(K)^\tau \}$, the bits of
$F(K)$ that correspond to possible detection events at Bob, has a $(\gamma =
max\{\frac{.75}{2}, \frac{1-\Delta}{2}\} )$ error rate.

\item Eve and Bob perform bit distillation. Eve sends Bob the set $\{\nu\} \oplus F(K')$,
where $F(K')$ is Eve's flawed version of $F(K)$. Bob then compares $\{\nu\} \oplus F(K')$ to
$\{z\}$ and sends Eve a list of which bits to include.  This set of events will be about
half of the events included by Eve and Alice.

\item Eve and Bob perform EC. Eve's error rate of $\{F(K')^\tau\}$ is $\gamma$, so her key
will have a $(\frac{\gamma}{2})$ error rate relative to Bob's key.

\item Bob will calculate a $.1675 \le (QBER = \frac{\gamma}{2} )\le .25$ and conclude that Eve
must be involved.

\end{enumerate}

\section{Conclusions}

The security of the shared-secret authentication lies in Eve's inability to predict the
secret bits, so it is imperative that the secret bits be well protected until Bob has a
chance to verify the sender's identity.  In each of the above protocols Alice leaks
information about $F(K)$ to the person with whom she is performing EC, so $F(K)$ is not
completely secure. However, as long as determining $K$ from $F(K)$ is a relatively
computationally-intensive process, then there is enough entropy in the shared secret during
the QKD session to prevent Eve from successfully carrying out a Man-in-the-Middle attack.

The significance of these protocols is that each of them could easily be implemented in
current QKD systems and would only require minor software modifications. Each of the
protocols can be used to authenticate the quantum channel of prepare-and-measure QKD
systems, such as BB84. However, note that in Protocol 2.b, Alice only has to know her basis
choice when she performs sifting and not when actually sending the photons. This feature
allows 2.b to actually be used for any 2-Basis QKD schemes that require bit distillation and
EC, even entanglement schemes. Similarly, Protocol 1.b only relies on Alice and Bob having a
bit stream with errors and a shared secret, implying that it can be used with all QKD
schemes, even no-switching QKD \cite{no-switch}, as long as the QKD schemes require EC.

Suppose that the roles in the sifting and EC were reversed, such that Bob's key prior to EC
was assumed to be correct, and Bob helped Alice correct her bits that differed from Bob's.
Alice would then calculate the QBER, and they could use the protocols for Alice to verify
that the photons were detected by someone who knows $F(K)$, Bob, and that she communicated
with him for the sifting and EC.  For example, Protocol 2.b would only differ in that Bob
would send Alice $\{z\}\oplus F(K)$ for sifting, and they would change roles for EC.
Therefore, Alice could also authenticate Bob's identity, and they could adjust the protocols
so that they can authenticate each other for every QKD session.

It should be noted that the protocols presented in this article belong to a more general
class of authentication protocols that use a shared secret, a symmetric-key algorithm, and
monitoring of the QBER to detect a Man-in-the-Middle attack. These four were chosen to
represent the versatitlity and utility of the protocols, but were certainly not inclusive of
all of the ways to use classical cryptography to authenticate a QKD quantum channel.
Alternative protocols could be created by replacing the $F(K)\oplus (Information)$ step by
running the information through an algorithm such as $AES$, or varying where the
encrypt/decrypt takes place, among other options. As was shown, the QBER induced by a
Man-in-the-Middle attack would vary between the different protocols, but many of the
possible protocols in this class are more than sufficient for authentication purposes.

I would like to thank my colleagues at LLL, LTS, and NSA for their insightful comments.
\end{multicols}

\begin{multicols}{2}

\end{multicols}

\end{document}